# A general cloak to shift the scattering of different objects

**Wei Li, Jianguo Guan[1], Wei Wang, Zhigang Sun, Zhengyi Fu**
E-mail: guanjg@whut.edu.cn

State Key Lab of Advanced Technology for Materials Synthesis and Processing, Wuhan University of Technology, Wuhan 430070, China

**Abstract.** By introducing a concept of "minifying layer", we propose a universal device to hide the position of an object by shifting its scattering without scaling effect for a distance, independent of the direction of incident wave and observation. Unlike the previously reported similar cloaks whose configurations are relevant to the objects to be cloaked. The design here is able to shift the scattering of various objects by one cloak. Moreover, via adjusting the parameter of the "minifying layer", tunable magnifying or even minifying effect on the shifted scattering can also be achieved. The performances of the devices are demonstrated by finite element simulations. This design provides a more flexible way to shift the scattering of an object, and thus can be expected to have wide applications.

Transformation optics, which allows for control over the electromagnetic (EM) wave and light wave at will, is attracting more and more research interests.[1-21] Due to the flexibility of the coordinate transformation method,[1, 7] many interesting devices such as superlenses,[2] superabsorbers,[3] superscatterers,[4] beam shifters and splitters, field rotators,[11] concentrators,[12] wormholes [13] and invisibility cloaks [5, 14-20] have been proposed based on metamaterials. Among them, the most typical and fascinating application is invisibility cloaks which render the objects invisible to the outside observers by bending the light. One of the important advantages of the transformation optics based invisibility cloaks is that the cloak configurations are irrelevant to the objects to be masked.[22] This feature makes the design of the cloaks independent of the shape and constituents of the cloaked objects, and provides a universal cloak for various objects. Inspired by the transformation optics, lots of valuable designs continuously come forth. Y. Lai et al. used a "complementary medium" to achieve "cloaking at a distance".[23] The complementary medium, which was originally proposed to make perfect lenses [24-26], has been proved to be a special transformation medium [27] and actually has been used to construct a number of transformation optics based devices. For example, with the assistance of the complementary

---

[1] Author to whom correspondence should be addressed. Tel.: +86-27-87218832. Fax: +86-27-87879468.

medium, Y. Luo et al. [6] presented a special device, which visually transforms an object from its original place to another place and simultaneously magnifies the object. However, the device fails to shift the scattering of objects without magnification effect. Since an illusion close to the real image is usually more realistic and thus more deceptive to the observer, the non-scaling visual shifting effect is potentially of more practical significance. Very recently, Y. Lai et al. [21] generalized an illusion device by combining the two concepts of the "complementary medium" and the "restoring medium". Indeed, one can use the illusion device to cancel the original scattering of the object and then restore a scattering at another place. The visual shifting without scaling effect is achieved when the restored scattering is the same as the canceled one. However, the illusion device does not apply to the case when the cloaked object changes the relative position with respect to the cloak. Besides, both of the two aforementioned illusion devices have a common disadvantage that the configuration are sensitive to the size, shape and constituents of the cloaked object. That is, when a new object needs to be cloaked, the cloak should be re-designed and/or rebuilt even when the new object is small enough to be put into the cloaked region. As a result, the design and fabrication of such a cloak would be complicated. To our best knowledge, so far no published works deal with the above problems despite their importance. In this paper, by introducing a "minifying layer", we demonstrate that a cloak, which is a universal cloak for various objects, visually shift the location of an object without changing the size of the virtual object. The cloaking effect is not affected by the relative position of the cloaked object and the cloak. Moreover, by controlling the constitutive parameters of the minifying layer, one can also make the shifted scattering to be magnified or even minified arbitrarily. Although the universality of the cloak is lost when the scaling effect is introduced, the minifying layer still provides a flexible manipulation to control the scaling effect of the shifted scattering. In the following, we will give the structure of the cloak and evaluate the properties of it by numerical simulations.

We start from the cloak exhibiting shifting and magnifying effect [6]. That device is actually based on a concept of "scaling device" [10] which is composed of two parts (see figure 1), one is the core part and the other is the shell part. If the background medium where the cloak is placed is homogeneous, the material parameters of the device can be given as [10]

$$\eta_r = \frac{r-B}{r}\eta_{bg}, \eta_\theta = \frac{r}{r-B}\eta_{bg}, \eta_z = \frac{r-B}{A^2 r}\eta_{bg} \tag{1a}$$

for the shell part and

$$\eta_r = \eta_\theta = \eta_{bg}, \eta_z = M^2 \eta_{bg} \tag{1b}$$

for the core part, with $A = (Ma-b)/(a-b)$ and $B = ab(1-M)/(a-b)$. $\eta_i$ ($i=r, \theta, z$) and $\eta_{bg}$ represent permittivity and permeability of the cloak and the background medium, respectively. $a$ and $b$ are the radii of the core and the entire cloak respectively, as indicated in figure 1(a). $M$ is a constant which is related to scaling and shifting effect. $M > 1$ for a magnifying device while $M < 1$ for a minifying device. As indicated in Ref. [6], by simply putting a PEC object into a hole in the core part of a magnifying device, the scattering of the PEC object can be shifted outside of the cloak with a magnification. Of course, not all the magnifying devices are suitable for shifting the scattering of an object. $M > b/a$ should be satisfied such that the PEC object put in the core can be visually shifted out of the cloak (figure 1(b)). Moreover, as we will point out later in this paper, the position where the object is placed also greatly affects the performance of the cloak.

For simplicity, we consider the TE polarized wave case in 2D in the following. In this case, only $\mu_r$, $\mu_\theta$ and $\varepsilon_z$ are related. If we presume that the cloak is placed at the origin of a polar coordinate system ($r$, $\theta$). The background medium where the cloak is placed is free space, which means $\mu_{bg} = \varepsilon_{bg} = 1$. Therefore, the materials in the core have the parameters of $\mu_r = \mu_\theta = 1$ and $\varepsilon_z = M^2$ according to Eq. (1b). For the outside observers, the permittivity of all the materials in the core is equivalently scaled by a factor of $1/M^2$, and every point ($r_c$, $\theta_c$) in the core is visually shifted to ($Mr_c$, $\theta_c$) [3-4, 6]. A cylindrical hole, with a radius of $c$ and a center at ($r_0$, $\theta_0$), is made in the core. When the hole is empty (filled with air or vacuum), the cloak is equivalent to a cylinder with parameters of $\mu = 1$ and $\varepsilon = 1/M^2$, a center at ($Mr_0$, $\theta_0$) and a radius of $Mc$. However, when a PEC cylinder is filled into the hole, it will be visually scaled and shifted as a bigger one with a center at ($Mr_0$, $\theta_0$) and a radius of $Mc$. If $Mr_0 < b$ or $Mr_0 < Mc$, obviously in these cases the shifted image intersect with the cloak. In this paper we only consider the case that the shifted image is totally out of the cloak. Therefore in the following discussion we presume that $Mr_0 > b$ and $r_0 > c$. If we define the distance between the virtual object and the device as the nearest distance of their outmost boundaries, that is $D = Mr_0 - b - Mc$. $D > 0$ means the shifted image is completely outside the cloak, and $D < 0$ means that the shifted image partially remains inside the cloak. We find that the distance $D$ can be an arbitrary positive value in principle just by controlling the parameter $M$, while other parameters, i.e., $b$, $c$ and $r_0$, also affect the shifting performance. The inequalities $M > b/(r_0-c)$, $Mr_0 > b$ and $r_0 > c$ should be satisfied such that the device can visually shift an object completely outside it. The scheme of such a device is plotted in figure 1(b). However, it is impossible to shift the scattering of an object without magnifying effect by using this device. Moreover, the farther we want to shift the cloaked object, the "larger" the identical object we will get. Besides, if the cloaked PEC cylinder is replaced by another PEC object with smaller size or a non-PEC object, the cloak will not work properly. Because the scaling effect is also applied to the material parameters of the cloaked object, any medium except for PEC in the cloaked region would be virtually changed when its position is being shifted and its size is being scaled. Therefore, this cloak only works to PEC objects which are just fitted into the cloaked region.

It would be more meaningful if a device visually shifts a cloaked object with exactly the same size or even intentionally control over the size of the shifted image. Moreover, when the total scaling factor is controlled to be 1, the applicable cloaking targets of the resulting cloak is not limited to the PEC objects with given shape and size. Various objects are expected to be cloaked by using the same device. In the following, some improvements are made to achieve this purpose. Notice that the virtual object is magnified by a factor of $M$ with respect to the cloaked object in figure 1(b). Thus we add a shell with an inner radius of $a'$ and an outer radius of $b'$ ($b' = c$) that surrounds the cloaked object and scales the object by a factor of $m = 1/M$ to counteract the magnifying effect while keeping the shifting ability of the device. The scaling on the parameters is also counteracted, such that any objects with various shapes or constitutive parameters can be shifted by one cloak. The scheme is illustrated in figure 1(c). The added shell, which has the ability of minifying the scattering of the cloaked object, is called the "minifying layer." The shifting mechanism can be explained in this way: at the first the combination of the cloaked object and the minifying layer at ($r_0$, $\theta_0$) is visually scaled by a factor of $M$ and shifted to $(Mr_0, \theta_0)$. At the new place, the minifying layer scales the cloaked object by a factor of $m$ ($m = 1/M$) but without shifting effect. Thus, the shifted combination can be further identical to a cylinder object the same as the cloaked one but at another place. The distance between the virtual image and the device can be written as $D = Mr_0 - b - a'$ (with $Mr_0 > b$ and $r_0 > a'$), and it can also approach infinity in mathematics by only increasing the parameter $M$. However, it may be difficult to get a big $D$ value with respect to the materials forming the device. To completely shift the virtual image out of the cloak, $M > (b + a')/r_0$, $Mr_0 > b$ and $r_0 > a'$ should be all satisfied. The general expression of the constitutive parameters of a scaling device with a center at the origin has been formerly given by Eq. (1). In order to express the material parameters of the minifying layer concisely, we define another polar coordinate system $(r', \theta')$ whose pole is at the geometric center of the minifying layer, i.e., $(r_0, \theta_0)$ in the old polar system $(r, \theta)$. With the new system, the material parameters of the minifying layer can be written as

$$\mu_r = \frac{r'-B'}{r'}\mu_{bg}, \mu_\theta = \frac{r'}{r'-B'}\mu_{bg}, \varepsilon_z = \frac{r'-B'}{A'^2 r'}M^2\varepsilon_{bg} \tag{2}$$

where $A' = (ma'-b')/(a'-b')$, $B' = a'b'(1-m)/(a'-b')$. $r'$ denotes the polar radius corresponding to a point at the minifying layer, with respect to the new polar system $(r', \theta')$. $a'$ and $b'$ are the inner and outer radii of the minifying layer, respectively. From the above description about the minifying layer, we have $b' = c$, and $0 < a' < c$. They decide the thickness of the minifying layer and the size of the cloaked region, as well as the constitutive parameters of the minifying layer by Eq. (2). $\varepsilon_{bg}$ and $\mu_{bg}$ are the material parameters of background medium where the cloak is placed. Usually, we wish the cloak work in air or vacuum, i.e., $\varepsilon_{bg} = \mu_{bg} = 1$. If let the scaling factor of the minifying layer $m = 1/M$, then the total scaling factor is $m \times M = 1$. The shifted virtual image should have the same size and constitutive parameters as the cloaked object. The scaling effect is eliminated. Therefore, different objects with different shapes, sizes and/or constitutes can be cloaked by the same device. By tuning the scaling factor of the minifying layer, the total scaling factor can be arbitrarily controlled. Thus the shifted virtual image can be arbitrarily scaled, either magnified or even minified. However, once the scaling effect is introduced, it goes back to the limitation of the cloak in Ref. [6], i.e., the cloak will no longer work properly for any object excepting the PEC objects with which the cloaked region is just filled up.

To demonstrate the properties of the cloak, we carry out the full wave simulations in TE polarized waves by using the finite elements method. In all cases the wavelength $\lambda$ in free space is 4 m. The parameters of the device we use in the simulations are $a = 3$ m, $b = 5$ m and $M = 5$. The parameters of the minifying layer which is located at (1, 1) (represented in Cartesian system, the same below) are $a' = 0.8$ m, $b' = 1.5$ m and $m = 0.2$. We firstly consider the case when the object to be cloaked is a PEC cylinder which is just fitted into the cloaked region. The equivalent PEC is thus calculated to be a cylinder with the same size and a center at (5, 5). Figure 2(a) and (b) show the total electric distributions induced by the effective bare PEC cylinder and the cloaked PEC cylinder, respectively, when the incident wave is a plane wave from left to right. The results show the good consistence of the two patterns outside the device. If the incident direction of the plane wave is from top to bottom, the coincidence of the two patterns is still kept (see figure 2(c) and (d)). This implies that the device can shift the scattering of the PEC cylinder effectively, without a scaling effect.

Another important characteristic which differentiates our cloak from the previous similar ones [6, 21] is that the cloaking effect of our cloak is independent of the constituents, shape, size (so long as it is small enough to put into the cloak), and position of the cloaked object. This advantage originates from that the total scaling factor of the cloak equals to 1. Here, to verify this characteristic, we apply the cloak used in Fig. 2 to shift the scattering of a non-PEC object such as a dielectric. By replacing the PEC object in figure 1 with a dielectric of $\varepsilon = 3$ and $\mu = 1$, figure 3(a) and (b) give the field distribution of the object with and without the cloak when a plane wave is incident from left to right. It can be seen that almost equivalent electric field distributions are obtained for the equivalent dielectric (figure 3(a)) and the cloaked dielectric (figure 3(b)).

Then a PEC object which is smaller in size than the one cloaked in Fig. 2 and of different shape is cloaked. We cut off the left part of the PEC object in figure 1(a) and re-perform the numerical calculation. Figure 4(a) shows the simulation result. For comparison, the field distribution of the PEC semicylinder without cloaking at the shifted place interacting with the same plane wave is also given in figure 4(b). Clearly, the field pattern outside the cloak in figure 4(a) is identical to that of the corresponding region in figure 4(b). This result indicates that the cloak is applicable for the objects with different sizes and shapes. On the other hand, as the PEC semicylinder can be viewed as a cylinder inhomogeneous object with air in the left part and PEC in the right part, the result also implies that inhomogeneous object can be cloaked with this device.

To verify that the cloak still works when the relative position between the cloaked object and the device is changed, the PEC semicylinder in figure 4(a) is rotated 90 degrees clockwise. The scattering pattern under the same plane wave irradiation is shown in figure 4(c). When it compares with figure 4(d), which is the scattering pattern of the object without the cloak and placed at the shifted place, there is almost no difference between them. Therefore, it is clear that the cloaking effect is not affected by the relative position between the cloaked object and the device.

The above numerical simulations show that the cloak can shift the scattering of various objects effectively. The flexibility of this configuration also makes it possible to magnify or minify the shifted scattering of the cloaked object by only adjusting the scaling factor of the minifying layer. As two examples, here we give the simulation results of the two cloaks which respectively magnify and minify the shifted scattering by 1.5 times. To magnify the shifted image by 1.5 times, we set the scaling factor of the minifying layer as $m = 1.5/M$. Thus the total scaling factor is $M(1.5/M) = 1.5$. A PEC cylinder with a radius of 0.8 m is cloaked by this device. When a plane wave is incident from left to right, the field distribution is shown in figure 5(a). For comparison, in figure 5(b), we give the field distribution when the same plane wave impinges on a PEC cylinder with a radius of 0.8×1.5 m placed at (5, 5). As we can see, the two cases show almost the same field distributions outside the cloak. The minifying effect on the shifted scattering has also been verified similarly by numerical simulations (see figure 5(c) and (d)). It is worth noting that a cloak designed to have a scaling ability never applies for cloaking objects with different shapes or constitutive materials, because the constitutive parameters are scaled simultaneously. As the cloak in Ref. [6], the cloaks with the scaling abilities here are also only applied to PEC objects with which the cloaked regions are just filled up. However, here the scaling effect can be arbitrarily controlled other than being restricted by the shifting distance.

From the above discussion, we conclude that all the numerical calculations consist well with our theoretical predictions. Noticeably, the scattering field distributions of the cloaked PEC cylinder are almost equivalent everywhere to that of the bare PEC cylinder outside the device. This implies that the shifting effect holds for all angles in the plane. In fact, since the cloak is designed in the framework of the transformation optics and the materials are angle-independent, the shifting effect of the cloak is inferentially irrelevant to the viewing angle. Comparing with the invisibility cloaks, the cloaked objects here can "see" the outside world other than "blind" like in an invisibility cloak, as the waves can enter the cloaked region of the device. Moreover, there is no singularity in the material parameters of our cloak, so it is more feasible to be implemented with regard to the materials aspect. However, the materials that build the cloak are still anisotropic and inhomogeneous like most present cloaks, making it difficult to implement the cloak. By using the effective medium theory, the inhomogeneous and anisotropic materials that constructing the cloak can be replaced by layered homogeneous and isotropic materials, and the cloaking performance is well kept.[28-30] Once implemented, this cloak can be used to hide the real position of an object and mislead the observer. For example, by cloaking a plane in such a device, a radar-guided missile may fail to attack that plane. Similar cloaks can be designed for visible light and sound waves, motivating more interesting applications.

In conclusion, we have designed a cloak that shifts the scattering of an object to a distance in the framework of transformation optics. We showed that one cloak designed here can function properly for various objects with different constitutive parameters and shapes, suggesting a broader application of our cloak than the previous similar ones which usually need customizing the configuration for every object. By adjusting the parameters of the minifying layer, we can also magnify or minify the shifted scattering at will, exhibiting the flexibility of the design method.

This work was supported by the National High-Technology Research and Development Program of China under Grant No. 2006AA03A209, the Young Teacher Grant from Fok Ying Tung Education Foundation under Grant No. 101049 and the Ministry of education of China under Grant No. PCSIRT0644.

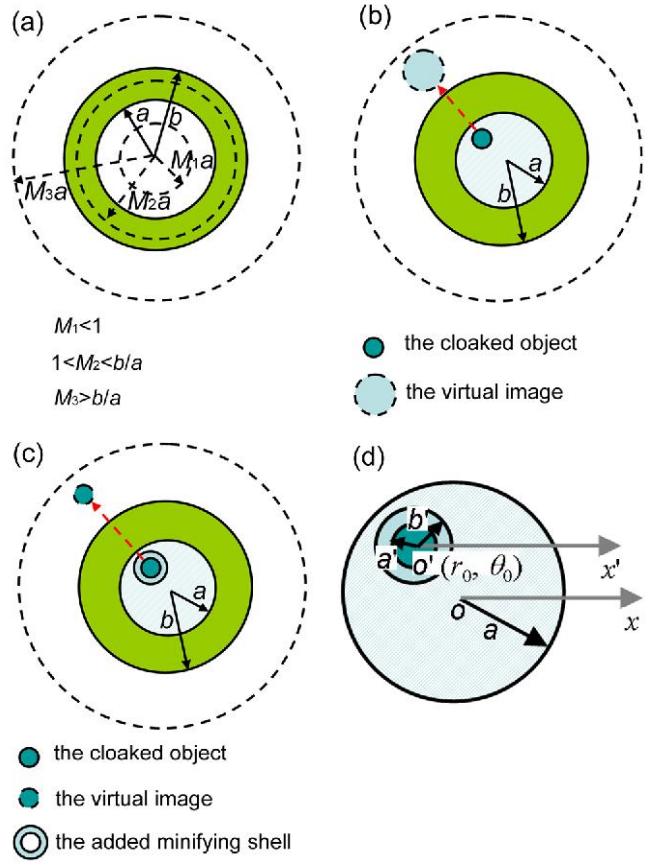

Figure 1. (a) The mappings transform a circular region $r<Ma$ ($M=M_1$, $M_2$, $M_3$) in virtual space to another circular region $r'<a$ in physical space and lead to two kinds of scaling device: minifying device ($M=M_1$) and magnifying device ($M=M_2$, $M_3$). (b) The scheme to visually magnify and shift an object. (c) The improved device with an added minifying layer visually shifts an object without changing its size. (d) The core part of the cloak in (c). It contains the minifying layer with a center at $O'$ ($r_0$, $\theta_0$), an inner radius of $a'$ and an outer radius of $b'$ ($b' = c$). Points $O$ and $O'$ are the poles of the polar systems ($r$, $\theta$) and ($r'$, $\theta'$), respectively. Geometries in virtual space are drawn in dashed lines. The dashed arrows in (b), (c) represent the visually shifting of the cloaked objects from their original places to another place.

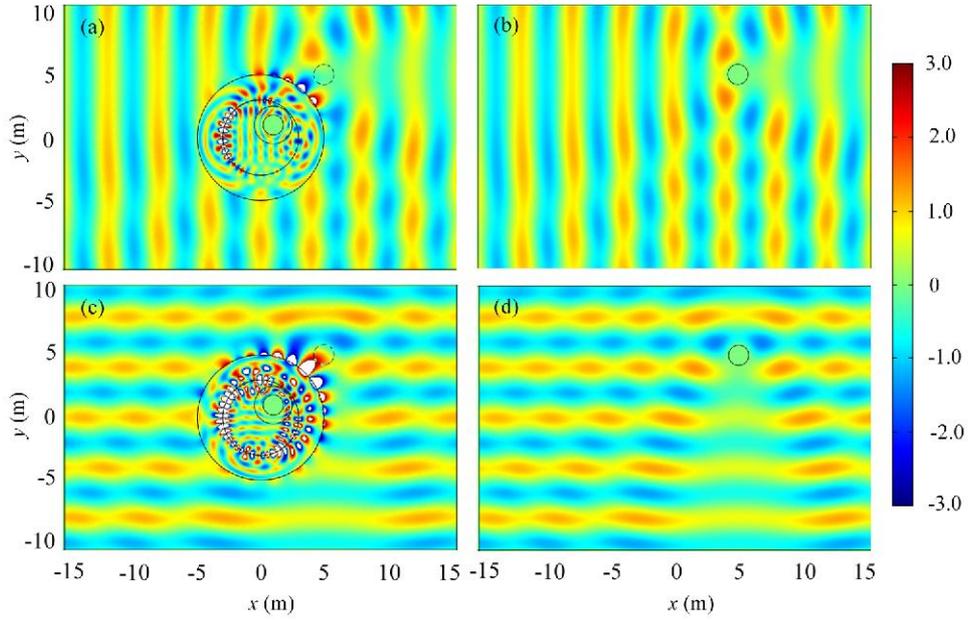

Figure 2. Electric field distribution of (a, c) a cloaked PEC cylinder with a center at (1, 1) and a radius of 0.8 m; (b, d) a naked PEC cylinder with a center at (5, 5) and a radius of 0.8 m under a plane wave illumination. The cloak is centered at (0, 0), and has parameters of $a$=3 m, $b$=5 m, $M$=5 and $m$=0.2. The minifying layer has an inner radius of $a'$ = 0.8 m, an outer radius of $b'$ = 1.5 m and a center at (1, 1). The dashed circles in (a, c) outline the boundaries of the virtual images. The plane wave is incident from left to right in (a, b) and from top to bottom in (b, d).

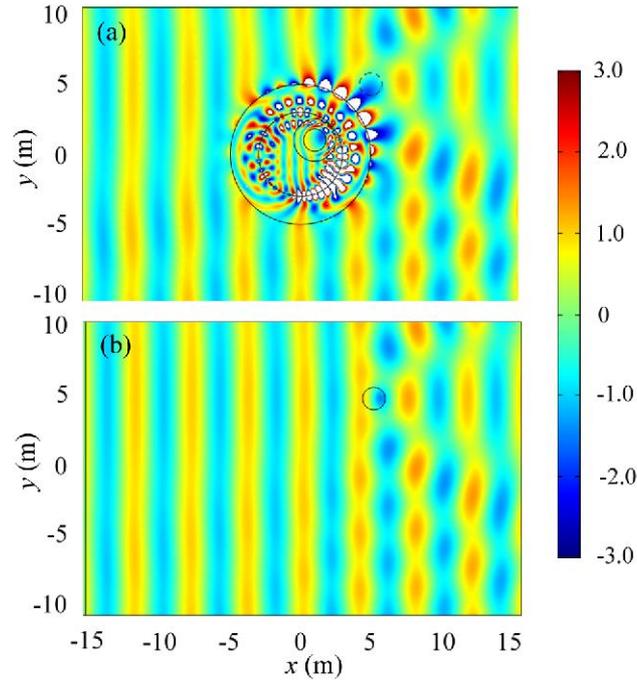

Figure 3. Electric field distribution of (a) a cloaked dielectric with parameters ε=3 and μ=1, a center at (1, 1) and a radius of 0.8 m; (b) an uncloaked dielectric with parameters ε=3 and μ=1, a center at (5, 5) and a radius of 0.8 m when a TE plane wave is incident from left to right. The cloak, which has parameters of $a$=3 m, $b$=5 m, $M$=5 and $m$=0.2, is placed at (0, 0), and the minifying layer has an inner radius of $a'$ = 0.8 m, an outer radius of $b'$ = 1.5 m and a center at (1, 1). The dashed circle in (a) outlines the boundary of the virtual image. The plane wave is incident from left to right.

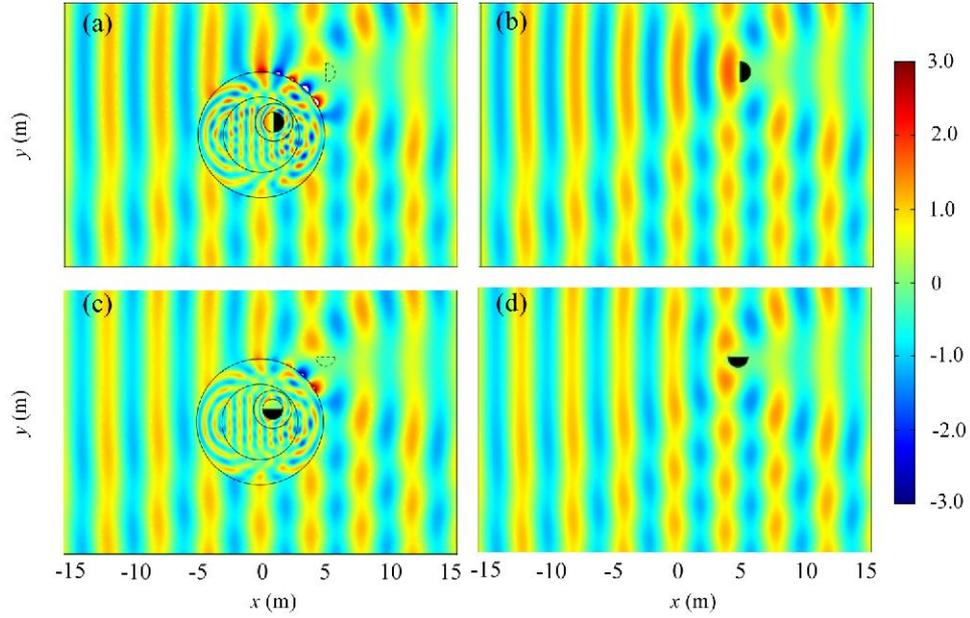

Figure 4. Electric field distribution of (a) a cloaked PEC semicylinder, (b) the same PEC semicylinder as in (a) without the cloak and located at the shifted place, (c) the PEC semicylinder in (a) rotated by 90° clockwise, and (d) the same PEC semicylinder in (c) without the cloak and located at the shifted place under a plane wave illumination. The cloak, which has parameters of $a$=3 m, $b$=5 m, $M$=5 and $m$=0.2, is placed at (0, 0), and the minifying layer has an inner radius of $a'$ = 0.8 m, an outer radius of $b'$ = 1.5 m and a center at (1, 1). The dashed semicircle in (a, c) outline the boundaries of the virtual images. The plane wave is incident from left to right. The PEC semicylinder is blackened for clarity.

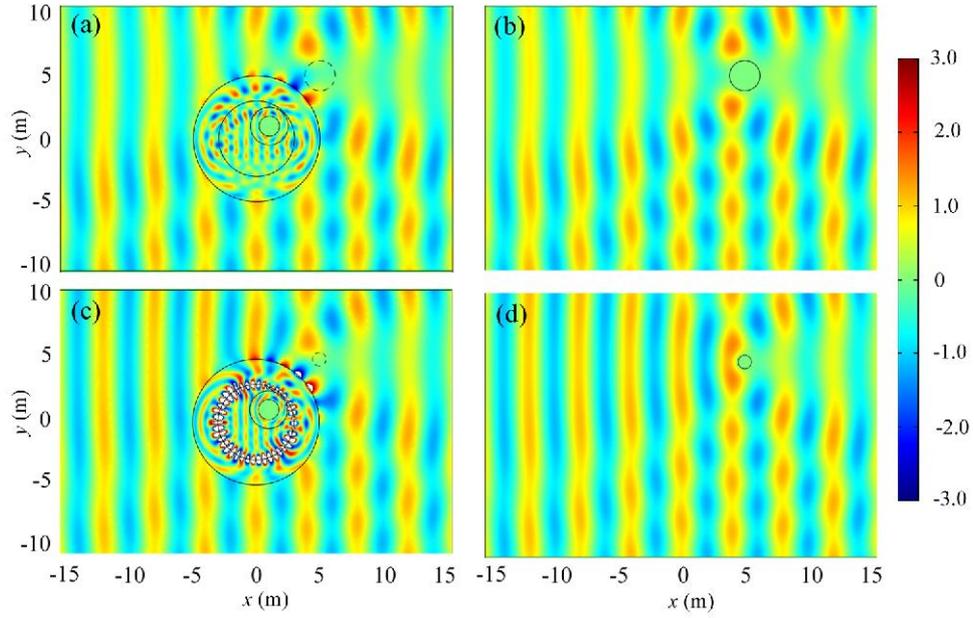

Figure 5. Electric field distribution of (a, c) a cloaked PEC cylinder with a center at (1, 1) and a radius of 0.8 m; (b) a naked PEC cylinder with a center at (5, 5) and a radius of 0.8×1.5 m;(d) a naked PEC cylinder with a center at (5, 5) and a radius of 0.8/1.5 m when a TE plane wave is incident from left to right. The cloak is centered at (0, 0), and has parameters of $a$=3 m, $b$=5 m, $a'$=0.8 m, $b'$=1.5 m, $M$=5 and $m$=1.5/5 in (a) [and $m$=1/(1.5×5) in (c)]. The dashed circles in (a, c) outline the boundaries of the virtual images.